\documentclass[a4paper,fleqn,usenatbib]{mnras}
\usepackage[T1]{fontenc}
\usepackage{ae,aecompl}

\usepackage{graphicx}	% Including figure files
\usepackage{amsmath}	% Advanced maths commands
\usepackage{amssymb}	% Extra maths symbols

\usepackage{color}
\usepackage{journals}

\title[Hot flow in SWIFT~J1753.5--0127]
      {Expanding hot flow in the black hole binary SWIFT~J1753.5--0127: evidence from optical timing}
 \author[Veledina et al.]
{Alexandra~Veledina$^{1,2}$\thanks{E-mail: alexandra.veledina@su.se}, Poshak Gandhi$^{3}$, Robert Hynes$^4$,  Jari J. E. Kajava$^{2,5}$,
  \newauthor  Sergey S. Tsygankov$^{2,6}$, Michail G. Revnivtsev$^6$, Martin Durant$^7$, and Juri~Poutanen$^{1,2}$ \\ 
$^1$Nordita, KTH Royal Institute of Technology and Stockholm University, Roslagstullsbacken 23, SE-10691 Stockholm, Sweden\\
$^2$Tuorla Observatory, University of Turku, V\"ais\"al\"antie 20, FI-21500 Piikki\"o, Finland\\
$^3$Department of Physics \& Astronomy, University of Southampton, Highfield, Southampton SO17 1BJ, UK \\
$^4$Department of Physics and Astronomy, Louisiana State University, Baton Rouge, LA 70803-4001, USA\\
$^5$European Space Astronomy Centre (ESA/ESAC), Science Operations Department, 28691 Villanueva de la Ca{\~n}ada, Madrid, Spain\\
$^6$Space Research Institute of the Russian Academy of Sciences, Profsoyuznaya Str. 84/32, Moscow 117997, Russia\\
$^7$Department of Medical Biophysics, Sunnybrook Hospital M6 623, 2075 Bayview Avenue, Toronto M4N 3M5, Canada\\
}

\begin{document}

\maketitle

\begin{abstract}
\noindent
We describe the evolution of optical and X-ray temporal
characteristics during the outburst decline of the black hole X-ray binary SWIFT~J1753.5--0127.
The optical/X-ray cross-correlation function demonstrates a single positive correlation at the outburst peak, then it has multiple dips and peaks during the 
decline stage, which are then replaced by the precognition dip plus peak structure in the outburst tail.
Power spectral densities and phase lags show a complex evolution,
revealing the presence of intrinsically connected optical and X-ray quasi-periodic oscillations.
For the first time, we quantitatively explain the evolution of these timing
properties during the entire outburst within one model, the essence of which is the 
expansion of the hot accretion flow towards the tail of the outburst.
The pivoting of the spectrum produced by synchrotron Comptonization in the hot flow is responsible for the appearance of the anti-correlation with the X-rays and for 
the optical quasi-periodic oscillations.
Our model reproduces well the cross-correlation and phase lag
spectrum during the decline stage, which could not be understood with any model proposed before.
\end{abstract}


\section{Introduction}

% (Fig. 1)
\begin{figure*}
\centering 
\includegraphics[width=13cm]{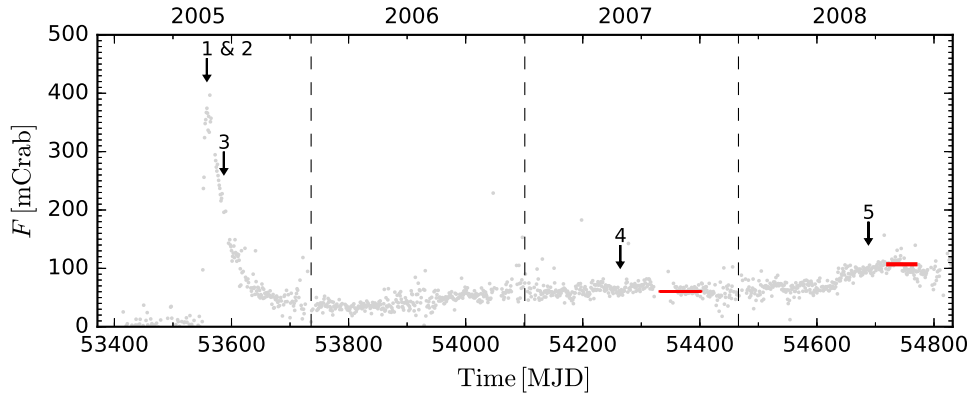}
\caption{
The 15--50 keV {\it Swift}/BAT light-curve of the BH binary SWIFT~J1753.5--0127.
The epochs considered in this work are shown with arrows.
Red stripes mark the {\it INTEGRAL}/ISGRI data used for spectral fitting (for epochs 4 and 5).
High-energy spectra for epochs 1, 2 and 3 are obtained from the {\it RXTE}/HEXTE data taken at the same time as {\it RXTE}/PCA.
}\label{fig:lc} 
\end{figure*}

The properties of accreting black holes (BHs) in X-ray binaries have been studied for almost half a century.
Most investigations relied on X-ray diagnostics. 
Spectra and their evolution, variability on the timescales as short as milliseconds and as long as months, and interconnection
of different energy bands allowed us to probe the regions in the immediate vicinity of the compact object \citep{ZG04,RM06,Gilfanov10,BS14,PV14}.
We know that the bulk of the X-rays are produced very close to the BH, probably, within $10R_{\rm S}$ (where $R_{\rm S}=2GM/c^2$ is the Schwarzschild radius and 
$M$ is the black hole mass), in an optically thin and hot medium
characterised by a Thomson optical depth of about unity and an electron temperature of $\sim$100~keV.
Some of these primary X-rays are reflected from the surface of the
cold thin disc producing characteristic features such as a fluorescent iron line and spectral hardening above 10~keV.
The location of the hot medium -- above the cold thin disc or within its truncated radius -- was debated for several decades \citep*{DGK07}. 
A related question concerns the source of seed photons for Comptonization which may come either from the cold disc \citep*[e.g.,][]{PKR97}
or from the synchrotron emission internally produced in the hot medium itself \citep{Esin97,PV09,MB09}.
Many observables suggest the cold disc is truncated far away from the black hole in the hard state, at about 100~$R_{\rm S}$: the characteristic suppression of 
the variability amplitude of the reflected emission compared to the primary X-rays \citep*{GCR00,RGC01}, the decrease of the iron line equivalent width with increasing 
Fourier frequency, X-ray luminosity and spectral hardness \citep*{RGC99,PFP15,BZ16}, the low temperature of the cold disc \citep{FZA01,CHM03} and 
reverberation lags \citep{DeMarco15reverber,DeMarco16reverber}.
Recently, yet another piece of evidence was presented.
This comes from the X-ray low-frequency quasi-periodic oscillations (QPOs), which were shown to be consistent with the Lense-Thirring precession of the 
hot medium by the detection of characteristic variations of the iron line blue/red shifts as the approaching or receding side of the cold disc is illuminated 
\citep{IvdK16}.
The flow precesses as a whole only under the condition of hot accretion \citep*{FB07,IDF09}, and its characteristic frequency depends on the outer radius of the
hot flow, which coincides with the truncation radius.

Because of the large amount of data and detailed studies of them, the X-ray information provided a reference point for studies in other wavelengths.
Whenever the variability in the optical, infrared or ultraviolet range
(hereafter, just optical) was investigated, it was compared to or
contrasted with the behaviour in the X-rays.
Three components can potentially contribute to the optical emission: the cold optically thick accretion disc, the hot optically thin X-ray emitting 
medium and an outflow/jet \citep[see review by][]{PV14}.
To break the spectral fitting degeneracy, timing information is used.
Simultaneous optical/X-ray data were obtained which, however, revealed another mystery: some observations showed a positive correlation with optical 
photons lagging the X-rays, consistent with simple reprocessing \citep{HOH98,HHC03,HBM09}, other demonstrated a very broad and nearly symmetric 
positive cross-correlation \citep{CMO10}, while in a number of cases a more complex structure containing a so-called precognition dip (anti-correlation) 
at negative lags (optical photons leading X-rays) was observed \citep{Motch83,Kanbach01,GMD08,GDD10,DGS08,DSG11}. 

In addition to the broadband variability, QPOs were identified in the optical power spectra \citep[e.g.,][]{Motch83,HHC03,DGS09,GDD10}.
They appear to share common frequencies with the X-ray QPOs for months of observations \citep{HHC03}, and it was recently shown that
they are phase connected \citep{VRD15,KCU16}. 

\begin{figure*}
\centering 
\includegraphics[width=13cm]{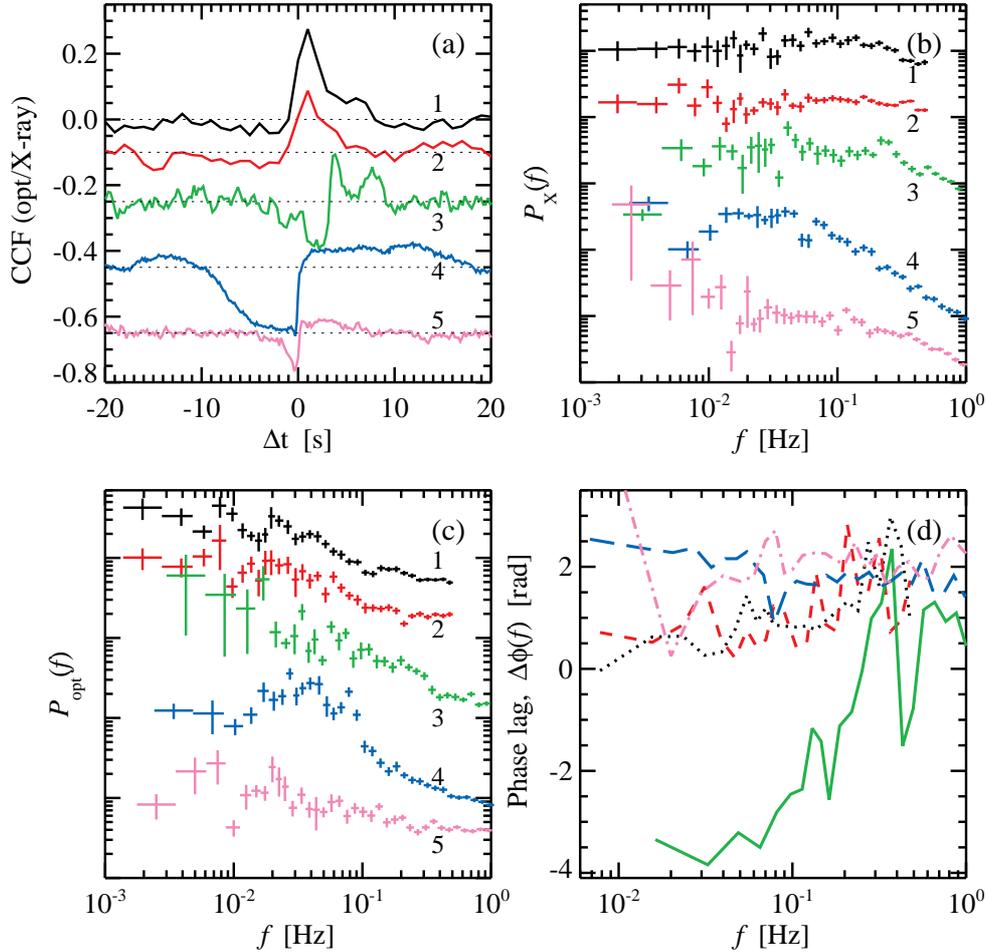}
\caption{
(a) Optical/X-ray CCFs (shifted vertically for clarity), arbitrary normalised (b) X-ray and (c) optical PSDs, (d) optical/X-ray phase lags from different outburst stages 
top to bottom: 
2005 July 6 (epoch 1, black lines, dotted line in panel d), July 7 (epoch 2, red, short-dashed), Aug 9 (epoch 3, green, solid), 2007 June 13 (epoch 4, blue, long-dashed), 
2008 Aug 10 (epoch 5, magenta, dot-dashed).
Positive time-lags in panel (a) correspond to lag of optical photons and dotted lines indicate zero level for each CCF.
We subtract $2\pi$ from the 2005 Aug 9 phase lag spectrum in panel (d) at frequencies below $\sim0.07$~Hz for clarity.
The evolution of CCF shape is apparent.
}\label{fig:ccf_psd} 
\end{figure*}

In this work we analyse several simultaneous optical/X-ray data sets
obtained for the BH binary SWIFT~J1753--0127.
The object underwent an outburst in the middle of 2005 and was active for over 11 years.
It is now returning to the quiescent state \citep{STB16Atel}.
The first four years of the outburst light-curve and the analysed epochs are shown in Fig.~\ref{fig:lc}.
This unique X-ray dataset revealed many interesting phenomena.
The temporal properties of the source, in particular, the correlation of the X-ray and optical light-curves, reveal dramatic changes throughout the 
outburst (see Fig.~\ref{fig:ccf_psd}, \citealt{HBM09,DGS08,DSG11}).
Recently, the spectral evolution throughout the ten years of the outburst was analysed \citep{KVT16}, with the aim of placing constraints on the accretion geometry and 
relevant radiative processes in this object.
The dependence of the ratio of the Comptonized to disc luminosity on
the spectral index revealed that the data points follow two different tracks above and below a critical flux.
At higher fluxes, the spectra are well explained by disc Comptonization, however, at lower fluxes, the inferred (observed) Comptonization luminosity is too high to be 
originating from upscattered disc photons.
These spectra are instead well reproduced by the Comptonization of synchrotron photons self-generated within the hot flow, 
implying a scenario where the cold disc recedes from the BH as the outburst proceeds.
Infrared/optical spectral energy distribution is well described by a power-law with $F_{\nu}\propto \nu^{1}$ \citep{FMF14,RTC15,Tomsick15}, 
which is consistent with the synchrotron emission from the stratified hot flow \citep{VPV13,PVR14}.
Interestingly, the object recently demonstrated several short incursions into the soft state at very low luminosities \citep{YYN15,SGA16}, challenging the simple one-to-one 
relation of the disc truncation radius and the X-ray luminosity.
The accretion history is likely an important factor in such relations.

In this paper we present the study of the evolution of the variability properties in SWIFT~J1753.5--0127.
We show that the temporal characteristics of the source can be explained by the expanding hot accretion flow, which emits in the optical through X-rays.
We investigate the evolution of power spectral densities (PSDs), cross-correlation functions (CCFs) and phase lag spectra and find that the optical emission can arise from simple 
reprocessing at the outburst peak, that it has an additional
synchrotron component in the outburst tail, and that the decline stage
can also be reproduced if the X-rays are produced 
by both synchrotron- and disc Comptonization.
Our findings support a picture where the cold accretion disc is receding towards the tail of the outburst, and a hot accretion flow is developing within its truncation radius. 

% (Fig. 2)

\section{Data}\label{sect:data_analys}

We reanalyse the optical data obtained on 2005 July 6 and 7 with the Argos CCD photometer on the McDonald Observatory 2.1-m telescope 
\citep*{NM04}, on 2005 Aug 9 with the FORS2 instrument on the VLT/UT2
telescope using HIT fast imaging mode \citep{OBrien08} and
on 2007 June 13 and 2008 Aug 10 with ULTRACAM \citep{Dhillon07} mounted on the VLT/UT3 telescope.
Simultaneous exposures were taken with \textit{Rossi X-ray Timing Explorer (RXTE)}/PCA in each case.
The X-ray data were analysed with the help of {\sc heasoft}.
Light-curves were created using the {\sc xronos} package.
The X-ray PSDs were normalised according to \citet{Leahy83} and the Poisson noise was subtracted.
The optical PSDs have arbitrary normalisation, no noise was subtracted from the data.

The data were published in \citet{HBM09} and \citet{DGS08,DGS09,DSG11}.
We briefly describe the data highlighting the features important for our modelling.
We further use the following notations: the outburst peak refers to 2005 July 6 and 7 (epochs 1 and 2 in Fig.~\ref{fig:lc}, respectively), the outburst decline (stage) denotes 
2005 Aug 9 (epoch 3),  and the outburst tail refers to 2007 June 13 and 2008 Aug 10 (epochs 4 and 5, respectively).
For the combined characteristics, such as CCF, cross-spectra and phase lags, we use simultaneous segments of data, 
and for PSDs we use longer light-curves with non-simultaneous segments, whenever possible 
\citep*[a detailed description of the calculation of these characteristics and their physical meaning can be found in][]{BP86,NWD99}.
The optical light-curves during the outburst tail were obtained by dividing, for each time-bin, the target flux by the flux of the comparison star.
For this reason, we use an arbitrary normalisation of the optical PSD.
In order to eliminate the long-term variations in the optical
light-curve, we subtract a linear trend.
This procedure, however, does not eliminate the contamination of the atmospheric noise, which is apparent in the optical PSDs.

The light-curves were split into segments; CCF, PSDs and cross-spectra were calculated in each of them, and then averaged over the segments. 
The cross-spectra were computed as the product of optical and conjugated X-ray Fourier images.
These are vectors on the complex plane, so we apply vector averaging procedure.
We plot the lengths of these vectors (denoted as CSD in figures and called just cross-spectra hereafter) and their angles on the 
complex plane (the phase lag spectra, $\Delta \phi$), which are determined in the interval $(-\pi,\pi]$.
The phase lags have $2\pi$ uncertainty.
Positive phase lags correspond to the delays of the optical light-curve with respect to the X-rays.
The cross-spectra show how the variability power of the correlated part of the signal is distributed over the Fourier frequencies $f$, while the phase lags show 
how the variability in one band is delayed with respect to another at each frequency.
Sometimes, the time lags ($\Delta t$) are a more intuitive representation of the delays, which are connected to phase lags through $\Delta t = \Delta \phi / 2\pi f$.
The errors are estimated using bootstrap method with $10^5$ trials.

\subsection{Outburst peak}

The simultaneous optical $V$ and X-ray data were taken during the outburst peak. 
The optical light-curves have 1~s time resolution.
The X-ray light-curves were obtained from Standard 1 mode with 0.125~s time resolution in the energy range 2--60~keV.
For each night, the data were split into 25 segments, 128~s long each.
The resulting characteristics are shown in Fig.~\ref{fig:ccf_psd} (in black and red for July 6 and 7, respectively).

The two X-ray PSDs are very similar and can be represented by a zero-centred Lorentzian (\citealt{NWD99,BPvdK02}).
The optical PSDs constitute a bump at frequencies below $\sim$0.1~Hz, and are flat at higher frequencies, which are likely dominated by the noise.
The CCF structure does not change within these two nights, however, its amplitude is somewhat reduced on July 7. 
The phase lags reveal a slowly increasing trend, again, nearly identical within two nights.
The CCF shape suggests there is one component in the optical which is positively correlated and delayed with respect to the X-rays,
and the delay time is in the range expected from thermal reprocessing \citep{HBM09}, making it a plausible source of optical variability.

\subsection{Outburst decline}

The $V$-band optical light-curve consists of several distinct segments.
We choose 11 segments which are simultaneous with the X-rays (3--20~keV), each lasting 64~s.
The time resolution is 0.25~s for the optical light-curve and 0.01~s for the X-rays.
The results are plotted in Fig.~\ref{fig:ccf_psd} (green lines).

The CCF during the decline stage consists of an anti-correlation dip at time lags between  $-$3~s and 3~s, followed by two peaks at $\sim$4 and 8~s.
Such CCF was not seen in any other object.
We calculate the CCFs for different X-ray energy ranges and find that they are nearly identical.
The X-ray PSD demonstrates a QPO at $\sim$0.25~Hz, but no apparent optical QPO can be seen.
The QPO period is the same as the separation of positive peaks seen in the CCF, so the second peak is likely arising from an aliasing, 
as was suggested in \citet{HBM09}.
In addition to that, we note a hump at $\sim0.6-0.7$~Hz (above the possible QPO harmonic), which was previously modelled as 
an additional broad QPO \citep{HBM09}.
The phase lag spectrum is a steep function of Fourier frequency.
It crosses the $\pi$ discontinuity point at least once, at $\sim$0.07~Hz, so we subtract $2\pi$ from phase lags for the sake of clarity.
Dramatic change of the CCF shape, now demonstrating an anti-correlation and double-peak structure at positive lags, suggests an alternative 
mechanism responsible for the optical emission, as simple disc reprocessing is no longer able to reproduce the observed CCF and phase lags.

\subsection{Outburst tail}

Two simultaneous optical (in filters $g'$ and $r'$) and X-ray (2--20~keV) datasets were obtained during the outburst tail.
We show the analysis for $r'$ filter, because the $g'$ demonstrates nearly identical behaviour.
The detailed analysis and modelling of the 2007 dataset is presented
by \citet{VRD15}; we repeat the main results here for the sake of completeness.
The simultaneous light-curves were split into 20 segments, each of about 146~s long.
Both light-curves have time resolution 0.143~s.
The resulting CCF, PSDs and phase lags are shown in Fig.~\ref{fig:ccf_psd} (blue lines).
Now the CCF has one pronounced dip at negative lag, followed by a pronounced peak at positive lags.
In addition to that, it demonstrates the oscillating behaviour which we attribute to the QPO at $f=0.08$~Hz, apparent in optical and in the X-ray PSD.
The phase lags are almost independent of frequency, but show a significant decrease at the QPO frequency.

The 2008 light-curves have a 0.195~s time resolution.
We use 24 segments, each of about 100~s.
The results are shown in Fig.~\ref{fig:ccf_psd} (magenta lines).
The CCF shows a single dip plus peak structure; no other structures can be seen.
Neither optical nor X-ray QPOs are apparent in the PSDs.
The amplitudes of the dip and the peak are somewhat lower, which might
be due to a larger contribution of atmospheric noise as compared to the previous case.
The white high-frequency ($\gtrsim$0.2~Hz) noise is also apparent in the optical PSD.
The phase lags increase with frequency, but show a kink at $\sim0.4$~Hz.

Both datasets obtained during the outburst tail are in agreement with a model where optical emission consists of two components: 
synchrotron emission from the hot accretion flow and the irradiated disc emission \citep*{VPV11}.
Whether this model can also account for the temporal properties during the decline stage is not immediately clear.
We further investigate this possibility below.

% (Fig. 3)
\begin{figure}
\centering 
\includegraphics[width=7cm]{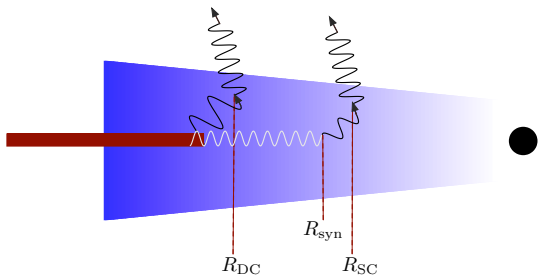}
\caption{
Sketch of the geometry considered.
The oscillations in white represent propagating fluctuations in the mass accretion rate, 
which are eventually transformed to radiation (black lines).
$R_{\rm DC}$ is the radius at which disc photons Comptonize, $R_{\rm syn}$ is the characteristic radius where the 
optical synchrotron photons are emitted and $R_{\rm SC}$ is the radius where synchrotron Comptonization proceeds.
}\label{fig:geometry} 
\end{figure}

\section{Model}\label{sect:model}

\subsection{General picture}

In this section we describe the model used for fitting of the computed characteristics \citep[see also][]{VPV11,VRD15}.
It is generally accepted that the X-rays are produced in the hot medium close to the BH, while the optical radiation often consists of the irradiated disc 
contribution together with synchrotron emission from the hot flow itself \citep{PV14}.
The X-ray broadband variability is driven by the fluctuations at a range of radii.
Each radius mainly contributes to one particular frequency, and the accumulated variability from larger radii propagate through the flow towards the compact 
object, thus the radiation produced at particular radius varies at frequencies up to the characteristic frequency of this radius \citep*{Lyub97,KCG01,CGR01,AU06,ID11}.

The X-ray emission is produced in the region of maximal energy release, within about 10~$R_{\rm S}$.
It is commonly accepted to originate from Compton up-scattering, but the source of seed photons is debated.
In the soft state, the bulk of soft photons for Comptonization is likely provided by the disc. 
In the hard state, the fraction of disc photons available for Comptonization is dramatically decreased because of the large truncation radius, so the 
synchrotron emission self-generated in the hot flow becomes an important, or even dominant, source of seed photons (\citealt{PKR97,Esin97,PV09,MB09}, \citealt*{VPV13}).
It is then likely that both sources coexist during the intermediate state \citep{V16}.

% (Fig. 4)
\begin{figure*}
\centering 
\includegraphics[width=14cm]{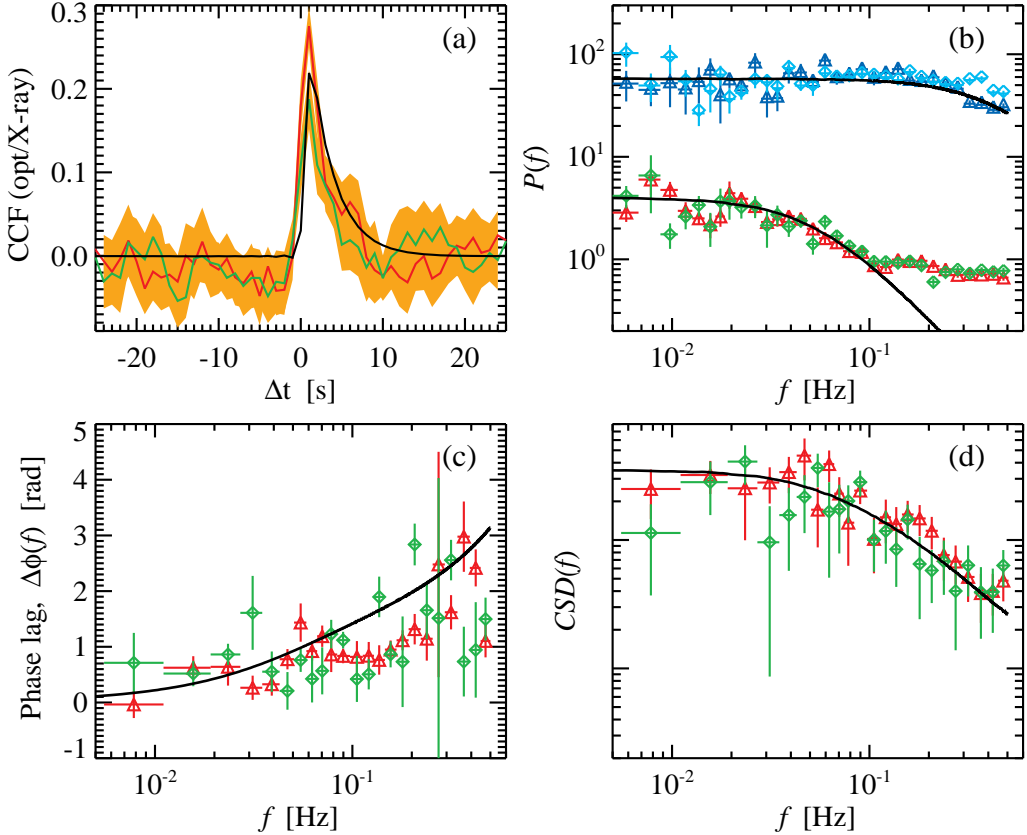}
\caption{
Characteristics of the light-curves observed during the outburst peak.
(a) Optical/X-ray CCF from 2005 July 6 (red) and July 7 (green). 
The yellow stripe represents the errors.
The peak on July 7 is reduced compared to that on July 6, however the difference is marginally significant.
(b) X-ray (top) and optical (bottom) PSDs from 2005 July 6 (blue and red triangles) and July 7 (cyan and green diamonds).
(c) phase lag spectra and (d) cross-spectra from 2005 July 6 (red triangles) and July 7 (green diamonds). 
The model is shown with a solid black line.
}\label{fig:2005_jul0607} 
\end{figure*}

In the propagating fluctuations model, the accretion rate changes first drive the disc Comptonization variability (close to the truncation radius) and then they propagate 
into the hot flow, exciting the X-ray fluctuations again, this time via synchrotron Comptonization (see Fig.~\ref{fig:geometry}).
We assume that the variations in accretion rate are directly mapped into
the broadband fluctuations of the disc- and synchrotron Comptonization continua.
These two mechanisms are independent of each other and so their luminosities are additive.
Moreover, the disc Comptonization likely proceeds close to the truncation radius, hence its light-curve lacks the high-frequency variations generated closer to the BH.
We simulate this by introducing a low-pass filter function with the Fourier image
\begin{equation}\label{eq:filter_func}
 H(f) = \frac{1}{(f/f_{\rm filt})^2+1}, 
\end{equation}
where $f_{\rm filt}$ is the characteristic frequency of fluctuation power damping, which is determined by the radius at which the disc photons are effectively up-scattered
(we use small letters with argument $t$ to denote variables in time domain and capital letters with argument $f$ for those in the frequency domain hereafter).
The disc Comptonization light-curve is obtained from the convolution of the mass accretion rate light-curve with the filter function.
The aperiodic synchrotron Comptonization light-curve is assumed to be proportional to the mass accretion rate light-curve.

The broadband variability is accompanied by the low-frequency QPOs, which can arise from the Lense-Thirring precession of the entire hot flow \citep{FB07,IDF09,IvdK16}.
We consider the model where the QPOs appear because of the anisotropy of the hot flow emission (\citealt*{VPI13}, Poutanen \& Veledina, in prep.).
The instantaneous emitted X-ray flux is determined by the current mass accretion rate, and to obtain the observed flux, we need to account for the angular distribution of 
radiation by multiplying the emitted power by the emission diagram (and also by the instantaneous solid angle occupied by the hot flow on the observers sky).
As the precession proceeds, the inclination angle of the flow changes resulting in the modulation of the observed X-ray flux.
Hence, the QPO modulates the accretion rate fluctuations in a multiplicative way.

The X-ray emission is reprocessed in the outer parts of the cool disc giving rise to the optical emission.
In addition to the disc, the hot flow synchrotron emission also contributes to optical wavelengths.
The optical light-curve is a weighted sum of these two components.

\subsection{Mathematical formulation}

The description we have introduced for the X-ray $x(t)$ and optical $o(t)$ light-curves can be expressed mathematically as
\begin{align}\label{eq:main}
 x(t)  & =  \varepsilon_{\rm m}\dot{m}(t+t_0)*h(t) + \left[1 + \dot{m}(t)\right]\left[1 + \varepsilon_{\rm x} q(t)\right] -1, \\
 o(t) & = s(t) + r_{\rm ds}d(t), \nonumber
\end{align}
where $\dot{m}(t)$ is the mass accretion rate light-curve, $h(t)$ is the filter function, $*$ sign denotes convolution, $q(t)$ describes the QPO light-curve, 
$s(t)$ and $d(t)$ are the synchrotron and irradiated disc light-curves, respectively.
All the light-curves have zero mean and represent deviations from the average quantity.
In the equation for the X-ray light-curve, the first term corresponds to the disc Comptonization and the term in brackets describes the emission coming 
from synchrotron Comptonization.
The parameter $\varepsilon_{\rm m}$ regulates their relative importance and $t_0$ is the delay between the disc and the synchrotron Comptonization processes.
The positive parameter $\varepsilon_{\rm x}$ gives the relative
importance of the QPO process and $r_{\rm ds}$ describes the relative importance of the disc and the 
synchrotron components.
The power spectra of both the synchrotron and disc light-curves are normalised so that their integrals are equal, thus 
$r_{\rm ds}$ gives the ratio of the total root mean square variability
amplitude of the disc to that of the synchrotron component.

% (Fig. 5)
\begin{figure*}
\centering 
\includegraphics[width=14cm]{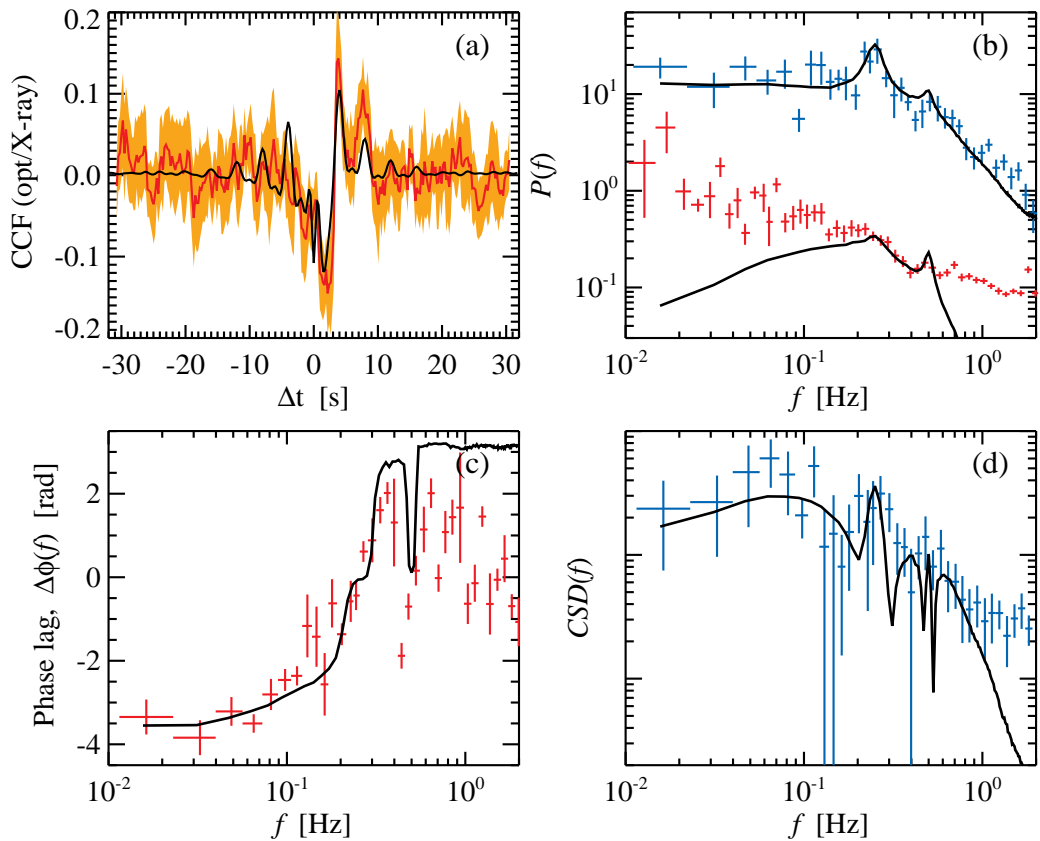}
\caption{
Characteristics of the light-curves observed during the decline stage of the outburst.
(a) Optical/X-ray CCF (red) and its errors (yellow), (b) X-ray (blue, top) and optical (red, bottom) PSDs, (c) phase lag spectrum and (d) cross-spectral density.
We subtract $2\pi$ from the phase lags at frequencies below $\sim0.07$~Hz for clarity.
The model is overplotted with a solid black line.
}\label{fig:2005_aug} 
\end{figure*}

The irradiated disc light-curve $d(t)$ is delayed and smeared with respect to the X-ray light-curve $x(t)$, which we simulate by a convolution of the 
X-ray light-curve with the disc response function, which we approximate by a simple exponential \citep{Pou02,VPV11}
\begin{equation}\label{eq:response}
  r(t) = \left\{
 \begin{array}{cc}
  \exp\left[ -(t-t_1)/t_2 \right]/t_2 , & t \geqslant t_1, \\
  0,                                             & t < t_1, 
 \end{array} \right.
\end{equation}
its shape is controlled by two parameters: the delay $t_1$ and the decay time $t_2$.
We note that the realistic transfer function should include a rather stable contribution of the disc and a variable contribution of the companion star, changing 
over an orbital period \citep[see, e.g. fig.~2 of][]{OBrHH02}.
We consider this generic response function as a rough approximation
having the least possible number of parameters.

The synchrotron radiation fluctuates in response to changing mass
accretion rates, however, in the opposite sense to the X-rays: 
when the mass accretion rate increases, the X-rays increase, but the synchrotron emission is suppressed because of the increased self-absorption. 
The optical emission is mostly produced at radii of about 15~$R_{\rm S}$ or further \citep{VPV13}, where the flow becomes effectively transparent to synchrotron.
These radii are somewhat larger than the characteristic region of major energy release, thus the synchrotron light-curve is expected to be smeared compared 
to the X-rays coming from synchrotron Comptonization, which we again simulate with the low-pass filter given by equation~(\ref{eq:filter_func}).
On the other hand, the region where optical photons are generated is likely to be situated close to the disc truncation radius, so the damping frequency for disc 
Comptonization is expected to be similar to that of synchrotron emission. 
We fix the damping frequency for synchrotron and disc Comptonization to be the same.
The characteristic damping frequency might differ substantially when the parameters of the hot flow change.
The synchrotron emission is modulated at the QPO frequency as the hot flow precesses \citep{VPI13}.
Similarly to the X-rays, the aperiodic synchrotron variability is multiplied by the QPO light-curve.

The quasi-periodic X-ray light-curve also modulates reprocessed emission at corresponding frequencies. 
However, the characteristic delay and profile shape depend on many unknown disc parameters \citep{VP15}, and in order to obtain the optical and X-ray QPOs 
in phase \citep[as observed in SWIFT~J1753.5--0127,][]{VRD15}, fine-tuning of parameters is needed.
We do not consider this possibility further.

% (Fig. 6)
\begin{figure*}
\centering 
\includegraphics[width=14cm]{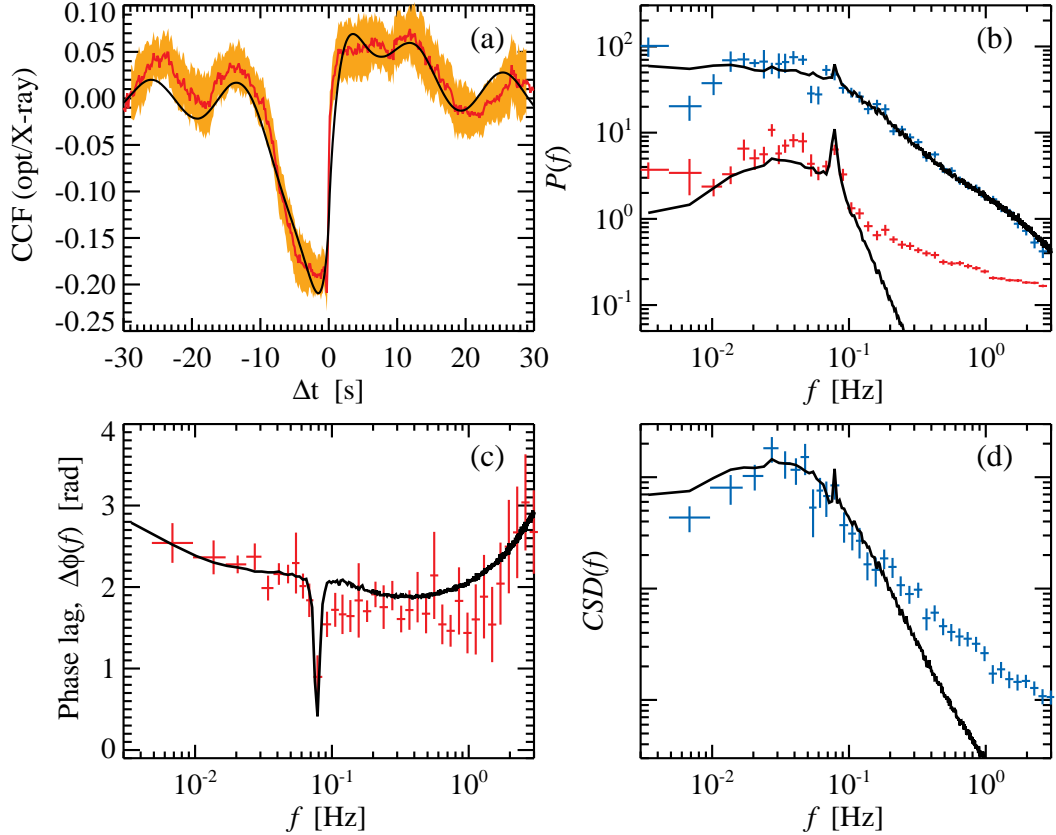}
\caption{
Characteristics of the light-curves observed during the outburst tail in 2007.
(a) Optical/X-ray CCF (red) and its errors (yellow), (b) X-ray (blue, top) and optical (red, bottom) PSDs, (c) phase lag spectrum and (d) cross-spectral density.
The model is overplotted with black solid line.
}\label{fig:2007} 
\end{figure*}

The mathematical description of the picture introduced is
\begin{align}\label{eq:opt}
 s(t) & = \left[1 - \dot{m}(t)*h(t)\right] \left[1 + \varepsilon_{\rm o} q(t)\right] -1, \\
 d(t) & = \left[ \varepsilon_{\rm m}\dot{m}(t+t_0)*h(t) + \dot{m}(t) \right] * r(t), \nonumber
\end{align}
where $\varepsilon_{\rm o}$ is a positive parameter responsible for the prominence of the QPO.
The first equation describes the synchrotron light-curve and is similar to the hot-flow X-ray light-curve (the term in brackets in equation~\ref{eq:main}), apart from the sign 
in front of mass accretion rate light-curve $\dot{m}(t)$ and the presence of filter function $h(t)$.
This minus sign denotes the anti-correlation of the synchrotron
emission and the local mass accretion rate fluctuations, which leads
to an anti-correlation of the hot flow synchrotron and X-ray 
variability.
The plus signs in front of $\varepsilon_{\rm x}$ (equation~\ref{eq:main}) and $\varepsilon_{\rm o}$ (equation~\ref{eq:opt}) reflect that the two QPOs come in phase.
The term in brackets in the second equation is just the X-ray
light-curve without the QPO.

% (Fig. 7)
\begin{figure*}
\centering 
\includegraphics[width=14cm]{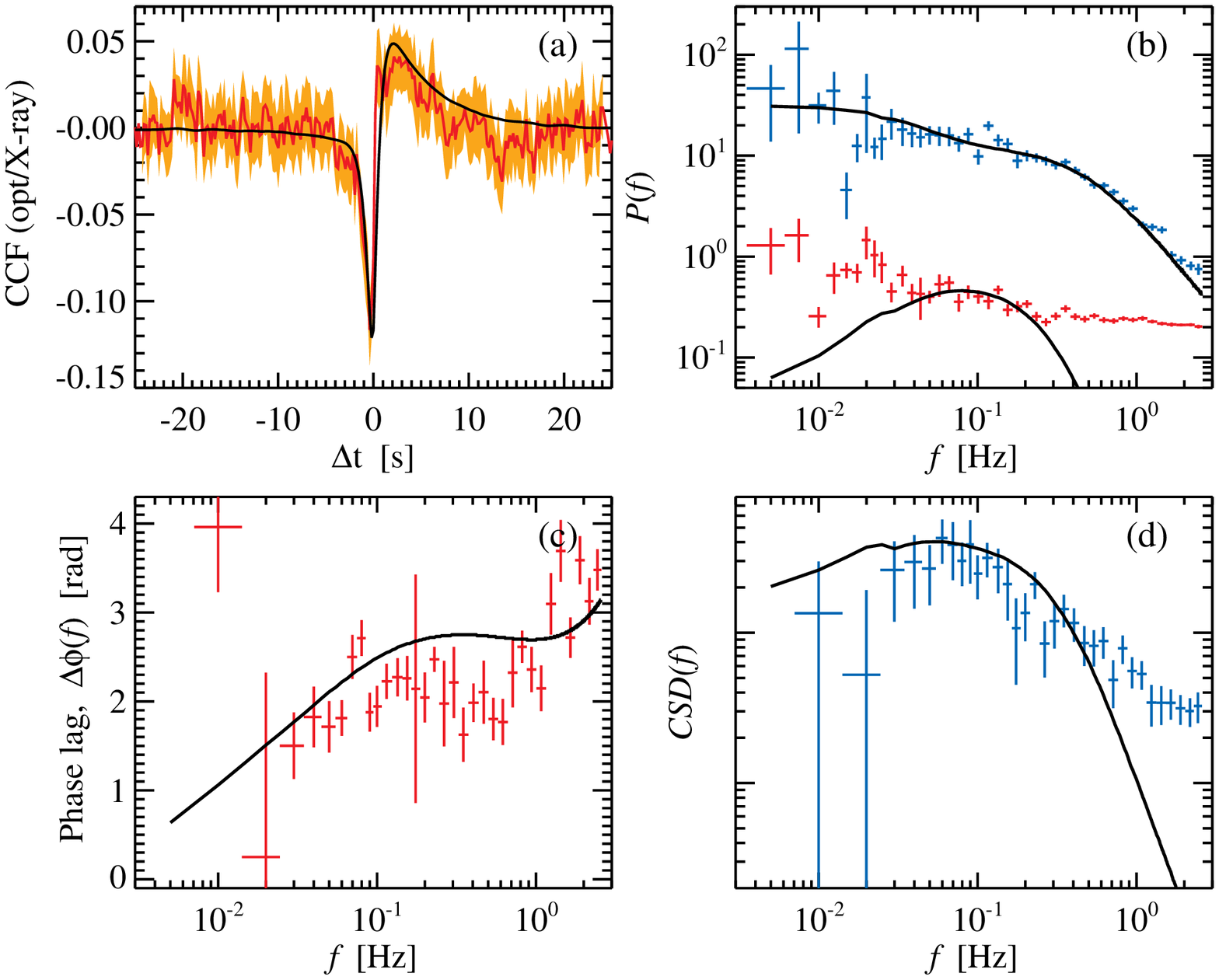}
\caption{
Characteristics of the light-curves observed during the outburst tail in 2008.
(a) Optical/X-ray CCF (red) and its errors (yellow), (b) X-ray (blue, top) and optical (red, bottom) PSDs, (c) optical/X-ray phase lag spectrum 
and (d) cross-spectral density.
The model is shown with a solid black line.
}\label{fig:2008} 
\end{figure*}

\subsection{Modelling procedure and study of parameter space}

The light-curve of the mass accretion rate as measured in the region of X-ray production, $\dot{m}(t)$, is calculated from the prescribed PSD 
using \citet{TK95} algorithm, with zero mean.
The PSD is described by a sum of Lorentzians of the form
\begin{equation}
 L_i(f)=\frac{r_i^2 \Delta f_i}{\pi [\Delta f_i^2 + (f-f_i)^2]}, \quad i=0,1, 2, 
\end{equation}
where $r_i$ describe their relative normalisations.
The broadband variability is represented by one or two zero-centred Lorentzians ($i=1,2$), thus $f_1=f_2=0$ \citep[after][]{NWD99}. 
The QPO and its second harmonic are described by Lorentzians ($i=0$) with central frequencies $f_0=f_{\rm QPO}$ and $f_0=2f_{\rm QPO}$, respectively.
We put the parameter $r_0=1$, so $\varepsilon_{\rm x}$ and $\varepsilon_{\rm o}$ determine the peak value of the Lorentzian.
Several methods to simulate the light-curve with the aperiodic noise and the QPO were proposed \citep{Burderi97,LS97,MBS03,ID12}.
We follow the latter approach and simulate the hot flow QPO from its PSD using the \citet{TK95} algorithm, with zero mean.

The disc to synchrotron ratio $r_{\rm ds}$ regulates the relative importance of the dip and the peak in the CCF. 
This parameter also affects the phase lags: for large $r_{\rm ds}$ the phase lags resemble those of the disc ($\Delta \phi$ increase with Fourier frequency), 
for smaller $r_{\rm ds}$ the phase lags resemble those of the synchrotron term ($\Delta \phi \sim \pi$).
The disc parameters $t_1$ and $t_2$ define the characteristic
frequency of suppression of high frequencies in the disc PSD and the frequency above 
which the phase lags arising from reprocessing start to substantially deviate from zero. 
In the CCF, $t_1$ is responsible for the shift of the positive peak and $t_2$ determines its width.
Parameter $f_{\rm filt}$ acts on the dip width in the CCF: a smaller
$f_{\rm filt}$ gives a wider dip. 
The prominence of bumps in the X-ray PSD is regulated by the parameter $\varepsilon_{\rm m}$, which also affects the relative amplitude of dips in the CCF.
Finally, $t_0$ determines the position of bumps in the X-ray PSD and
the shift of the dip in the CCF (with respect to zero).

The parameters $\varepsilon_{\rm x}$ and $\varepsilon_{\rm o}$ regulate the QPO prominence in the PSDs, the wave amplitude in the CCF
and the dip depth, as well as affecting the phase lags at $f_{\rm QPO}$. 
The QPO Lorentzian width $\Delta f_{\rm QPO}$ determines the characteristic timescale of coherent oscillations.

\section{Results of modelling}

The resulting CCF, X-ray and optical PSDs, phase lags and cross spectra are shown in Figs~\ref{fig:2005_jul0607}--\ref{fig:2008}
and the model parameters for equations~(\ref{eq:main}) and (\ref{eq:opt}) are listed in Table~\ref{tab:par}.
The first six parameters ($r_{\rm ds}$, $t_1$, $t_2$, $f_{\rm filt}$, $t_0$ and $\varepsilon_{\rm m}$) describe the broadband variability. 
The following four parameters ($\varepsilon_{\rm x}$, $\varepsilon_{\rm x}$, $\Delta f_{0}$, $r_{0}$) describe the appearance of the QPO, manifesting 
itself as waves in the CCF, narrow spikes in the PSDs, and the phase
lags at the QPO's fundamental and harmonic frequencies.

At the peak of the outburst we see no QPOs and therefore we set $\varepsilon_{\rm x}=0$.
Furthermore, there is no indication for a precognition dip. In the
context of the model this suggests a minor contribution of synchrotron emission
and thus $r_{\rm ds}\gg1$.
The X-rays are dominated by the disc Comptonization, hence we take $\varepsilon_{\rm m}\gg 1$ and $t_0=0$.
Finally the filter function can be omitted.
Under these conditions, the equations for the X-ray and optical light-curves reduce to $x(t)=\dot{m}(t)$, $o(t)=\dot{m}(t)*r(t)$.
We describe both the July 6 and 7 data with one set of model parameters.
The model for outburst peak is shown in Fig.~\ref{fig:2005_jul0607}.

The CCFs are well reproduced in the simple model of disc reprocessing.
The same conclusion was also reached by \citet{HBM09}, where more realistic disc transfer functions \citep{OBrHH02} were used to model the CCF.
The phase lag spectrum demonstrates a slowly increasing function of frequency and is in agreement with model predictions. 
The optical PSD is reproduced up to $f\sim0.1$~Hz, above which the model significantly underestimates the data.
We attribute this inconsistency to the presence of high-frequency atmospheric noise.
We note that the cross spectrum is well reproduced up to Nyquist frequency, suggesting that the model reproduces the entire correlated signal.

\begin{table}
\caption{Parameters of numerical modelling. 
}\label{tab:par}
  \begin{center}
\begin{tabular}{ccccc}
\hline
{Parameter}		  & 2005 July 6, 7 & 2005 Aug	& 2007	& 2008		\\
\hline 
$r_{\rm ds}$		   &$\infty$	&0.4		& 0.8		& 0.8	\\
$t_1$~(s)			   & 0.15           	&0.15	& 0.15	& 0.15	\\
$t_2$~(s)			   & 3.0          	&3.0		& 4.5		& 3.0		\\
$f_{\rm filt}$~(Hz)	   & -    	        &0.45	& 0.05	& 0.3	\\
$t_0$~(s)		  	   & -         		&1.8		& -		& -  \\
$\varepsilon_{\rm m}$  & -          	&1.8		& - 		& -	\\
$\varepsilon_{\rm x}$   & -         	 	&1.8, 0.5	& 0.4		& -	\\
$\varepsilon_{\rm o}$   & -          	&0.1,0.13	& 0.4		& -	\\
$f_0$~(Hz)		   & -        	  	&0.25	& 0.078	& - \\
$\Delta f_0$~(Hz)  	   & -         	 	&0.03	& 0.003	& -  \\
\hline
\multicolumn{5}{c}{X-ray broadband noise parameters}  \\
$\Delta f_1$~(Hz)	  & 0.46		&0.3		& 0.095 	& 0.03	\\
$r_1$			  & 1.0		&1.0 		& 1.3		& 0.5\\
$\Delta f_2$~(Hz)	  & -			& -		& 1.5		& 0.5\\
$r_2$			  & -			& -		& 0.9		& 0.15\\
\hline
     \end{tabular}
  \end{center}
\end{table}

The outburst decline stage is described by the equations~(\ref{eq:main}) and (\ref{eq:opt}) with all parameters being non-zero.
There are simultaneously two components in both the optical and the X-ray wavelengths.
The CCF demonstrates strong coupling of the broadband spectral variability to the QPO (Fig.~\ref{fig:2005_aug}), which now has two harmonics.
Both positive peaks are produced by the QPO.
There are two dips in the CCF: one at time lag $\sim-1$~s and another at $\sim2$~s.
They are produced by the anti-correlation of the synchrotron optical emission with the two X-ray components.
Although we did not introduce any time-delays between the synchrotron and its Comptonization, the dip appears at somewhat negative lags due 
to a significant contribution of the (positively correlated)
irradiated disc emission at $\Delta t \sim 0$, partially cancelling
the negative contribution of the synchrotron emission at zero lag.

The X-ray model PSD demonstrates a peak at 0.6~Hz arising from the interference of the two X-ray terms (Veledina 2016).
The optical model PSD is again somewhat different from the observed one at high frequencies, but the cross spectrum is well reproduced up to $\sim1$~Hz.
The rapid increase of phase lags with frequency appears in our model (mainly) because of the delay between the optical synchrotron light-curve with respect 
to the X-ray light-curve arising from disc Comptonization.
The increase of phase lags becomes somewhat more rapid when we include the QPO.
A sharp drop of phase lags seen at the double QPO frequency is due to the presence of the second QPO harmonic, which may arise from the anisotropy 
of the synchrotron radiation.

The 2007 dataset shows no indication of multiple dips and peaks in the CCF.
We interpret this as an indication of one component contributing to X-rays, namely, the synchrotron Comptonization, and put $\varepsilon_{\rm m}=0$.
The model has the first QPO harmonic revealing itself in both the CCF and in the phase lags (Fig.~\ref{fig:2007}).
We see that the phase lags are almost independent of frequency, as explained by the joint contribution of the synchrotron (having $\Delta\phi\sim\pi$) 
and the disc terms ($\Delta\phi\sim 0$ in the frequency range of interest).
At the QPO frequency the phase lags demonstrate a sharp dip, as the X-ray and optical QPOs come into phase.
The cross-spectrum is reproduced up to $\sim$0.2~Hz.

The 2008 dataset is described by the model with $\varepsilon_{\rm m}=0$ (only synchrotron Comptonization contributing to X-rays), 
and without QPOs, $\varepsilon_{\rm x}=0$, $\varepsilon_{\rm o}=0$, see Fig.~\ref{fig:2008}.
The phase lags demonstrate a kink at $\sim0.4$~Hz. 
Noting that our model has a characteristic frequency of synchrotron emission suppression $f_{\rm filt}=0.3$~Hz, we interpret this kink as a
transition from a phase lag spectrum dominated by synchrotron (for
$f\lesssim0.3$~Hz) to one dominated by the disc ($f\gtrsim0.4$~Hz).
We again see that the optical PSD and CCF demonstrate strong noise, hence we use the cross-spectral density spectrum to check that the correlated (intrinsic)
signal is well reproduced.
The cross-spectrum is reproduced up to $f\sim0.7$~Hz.

% (Fig. 8)
\begin{figure}
\centering 
\includegraphics[width=7cm]{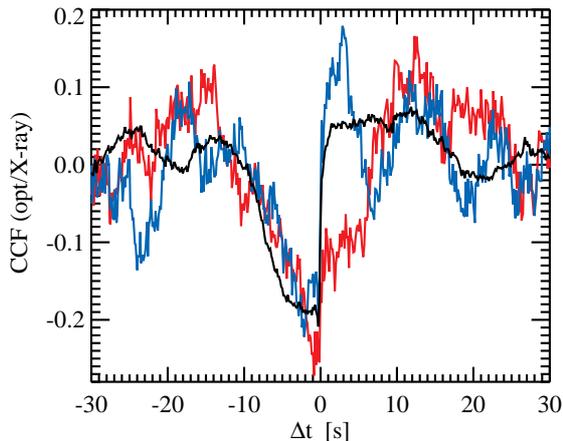}
\caption{
CCFs from two segments of 2007 data (red, blue) and the average CCF (black).
Variations of the shape are apparent, in particular, at small positive lags.
}\label{fig:ccf_change} 
\end{figure}

\section{Discussion}

\subsection{Deviations of the model from the data}

We find deviations of the model from the data in the optical PSDs at high frequencies (above about 0.5~Hz) in all datasets. 
We suggest these inconsistencies are mostly due to atmospheric noise.
Scintillation due to refraction from changing turbulent cells in the
atmosphere contributes to higher Fourier frequencies, complementing
the contribution of Poisson noise here.
Differential extinction variations may contribute to noise at lower frequencies (these effects are likely seen at low frequencies of the optical PSD from the outburst decline).

Considering the good agreement between the observed and modelled CCFs and phase-lags, the deviations of the cross-spectra are more puzzling. 
We find that the excess power at high frequencies can be accounted for by a more sophisticated shape of the filter function for synchrotron radiation 
(which determines the PSD shape at high frequencies).
For example, by adding a constant to the filter (equation~\ref{eq:filter_func}) we can achieve good agreement between the modelled and observed cross spectra.
This additional constant accounts for the fact that the synchrotron radiation is likely distributed within the flow, partially coming from its outer parts (and thus smeared) and
partially originating from the regions where the bulk of the X-rays are emitted.
The CCF is also somewhat better described in this case, as now we are able to account for the small negative spike seen at time lags close to zero in Fig.~\ref{fig:2007}a.
However, the model phase lags then deviate somewhat more from the data, suggesting we need a more realistic shape for the response function.

% (Fig. 9)
\begin{figure}
\centering 
\includegraphics[width=7cm]{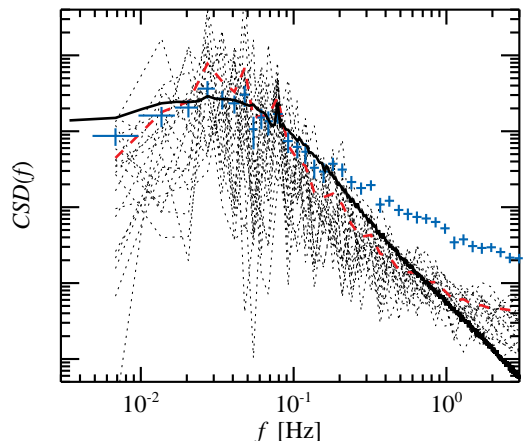}
\caption{
Cross-spectra from the 2007 data, calculated by the vector averaging of CSDs from different segments (blue crosses, same as in Fig.~\ref{fig:2007}, but scaled up by a factor of 20) 
and by averaging of CSD amplitudes of different segments (red dashed line).
The latter describes a typical CSD from one segment.
Cross-spectra from each segment are shown with black dotted lines.
The model (black solid line) describes the typical CSD better than the one obtained from vector averaging.
}\label{fig:csd_change} 
\end{figure}

Another reason for the discrepancy might be that we have implicitly assumed that the parameters do not change within one epoch.
However, the observed CCF shows large shape variations even during one observational run (see Fig.~\ref{fig:ccf_change}).
Each segment CCF can be reproduced by the model, but requires different parameters. 
Primarily, we expect the disc to synchrotron ratio $r_{\rm ds}$ to change.
We note that the (model) CCFs obtained from averaging of CCFs computed with different parameters is not equivalent to the CCF calculated from one parameter set, 
even if the parameters are the average ones.

To investigate the reason for the appearance of the high-frequency power in cross-spectra we checked their shape in every segment.
In Fig.~\ref{fig:csd_change} we show the cross-spectrum obtained from vector averaging, as before (blue crosses), and each segment CSD (black dotted lines).
We see that their shapes differ substantially at high frequencies.
This discrepancy is caused by the reduction of coherence at low frequencies, compared to higher ones and is seen as larger spread (at lower frequencies) 
between values from different segments (the two to three orders of magnitude spread at lower frequencies is opposed to an order of magnitude spread at 
higher frequencies).
Hence, the vector averaging effectively reduces the power at lower frequencies more, as compared to higher frequencies.

% (Fig. 10)
\begin{figure*}
\centering 
\includegraphics[width=14cm]{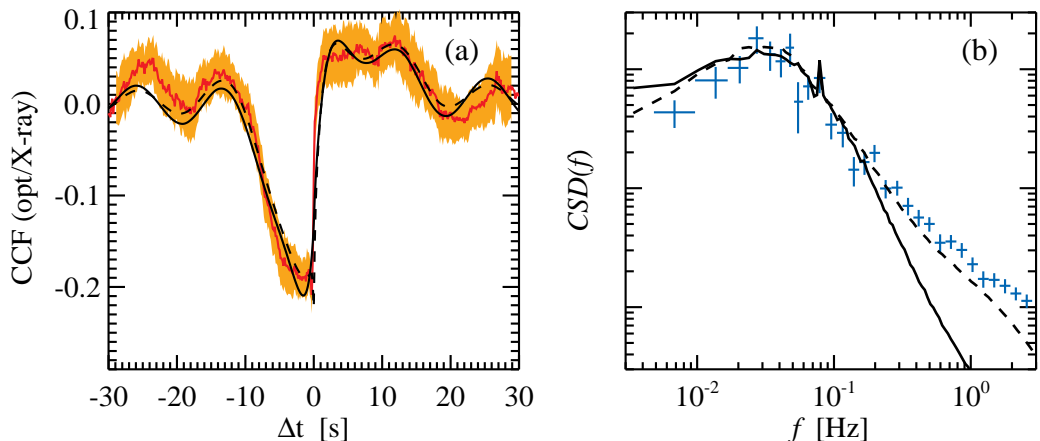}
\caption{
CCF (a) and CSDs (b) of the 2007 dataset, as in Fig.~\ref{fig:2007}. 
Black solid lines correspond to the model with constant parameters, dashed lines correspond to the model with varying disc to synchrotron ratio ($r_{\rm ds}$ parameter), all 
other parameters are fixed and same as for the solid line. 
}\label{fig:var_par} 
\end{figure*}

From the model point of view, the large spread at lower frequencies is likely to be the result of varying disc to synchrotron ratio within one observation.
Because the interplay of the two components is most prominent at lower frequencies, the spreading is higher here.
In fact, the change of the CCF shape (Fig.~\ref{fig:ccf_change}) is a manifestation of the same phenomenon as the reduction of cross-spectral power at low frequencies.
We note that the model cross-spectrum (black solid line in Fig.~\ref{fig:csd_change}) reproduces the shape of the cross-spectrum of a typical segment (red dashed line) 
substantially better than the one obtained from vector averaging.

To further investigate this hypothesis we perform modelling allowing the disc to synchrotron ratio change.
We assume that the $r_{\rm ds}$ takes values between 0 and 1.6 with flat probability distribution, while other parameters are fixed to values from Table~\ref{tab:par}.
For each value, we simulate the optical light-curve, compute CCF and CDS and then average these values over 1500 different realisations.
The resulting characteristics are compared to those modelled for constant parameters and with the data in Fig.~\ref{fig:var_par}.
We find that the shape of the cross spectrum has dramatically changed, now exhibiting the missing power at higher frequencies.
Moreover, the CCF now exhibits a small spike at negative lags,  similar to that observed.

We note that the standard procedure of improving signal-to-noise by averaging over a number of data segments \citep{vdK89} assumes that all segments have 
the same statistical properties, i.e. the time series is stationary.
We see that this assumption is violated in the cases we are considering (Fig.~\ref{fig:ccf_change}), warning similar problems for optical data in other systems.
Further data processing should take this limitation into account.

\subsection{Evolution of parameters}

Some restrictions on the parameters can be imposed from physical arguments.
The reprocessed disc emission is mostly coming from the outer, flared rim of the disc, thus we expect $t_1\gtrsim R_{\rm d}(1-\sin{i})/c$ 
(where $R_{\rm d}$ is the disc outer radius, \citealt{Pou02}).
For $i\lesssim75$\degr  \citep[the maximal inclination observed for a BH binary,][]{CJ14} we get $t_1\gtrsim0.07$~s.
The decay time $t_2$ is connected to the characteristic size of the disc.
Taking the semi-major axis $a=10^{11}$~cm \citep{NVP14}, we expect a
maximal outer disc radius $R_{\rm d}=6\times10^{10}$~cm (equal to the tidal truncation radius), 
constraining $t_2\sim2$~s.
The decay time obtained for the 2007 dataset is somewhat larger. 
The reason for that might be the increase of the delays from the companion star due to a change of the orbital phase (see fig.~2 of \citealt{OBrHH02}), 
which is completely ignored in our model.

\begin{figure}
\centering 
\includegraphics[width=8.5cm]{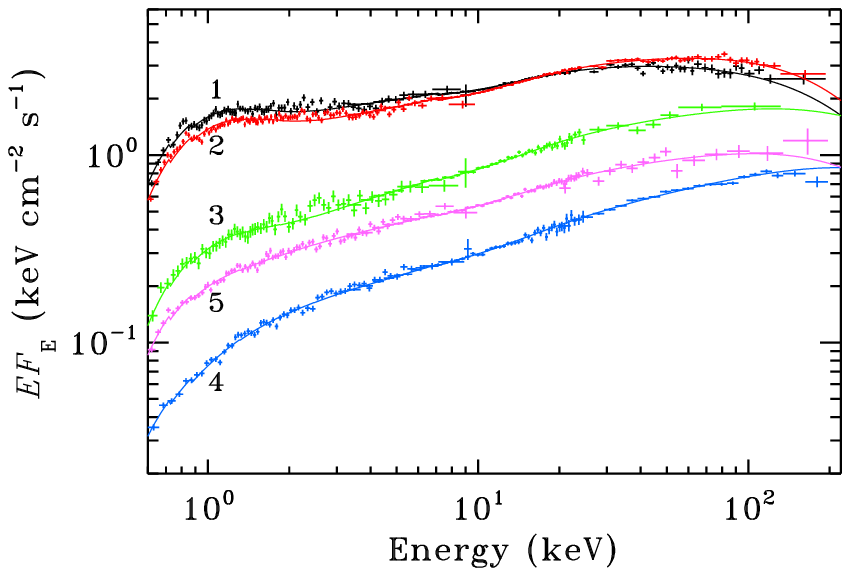}
\caption{
X-ray spectra of the considered epochs: 2005 July 6 (1, black), July 7 (2, red), Aug 9 (3, green), 2007 June 13 (4, blue), 2008 Aug 10 (5, magenta).
The spectra 1 and 2 are fitted using {\sc tbabs$\times$(diskbb+compps)} \citep*[respectively]{WAMC00,M84,PS96}. 
For spectra 3--5, only {\sc tbabs$\times$compps} was used. 
The data are from {\it Swift}/XRT, {\it RXTE}/PCA and {\it RXTE}/HEXTE (spectra 1, 2, 3) or {\it INTEGRAL}/ISGRI (spectra 4 and 5).
}\label{fig:spectra} 
\end{figure}

To explain the evolution of the temporal properties, we utilise the
truncated disc scenario and consider the appearance of an additional optical (synchrotron) component 
towards the tail of the outburst.
In our model, the truncation radius is connected to the synchrotron
filtering frequency, which is found to have a minimal value in 2007.
If they are not a result of noise contamination, our findings suggest that the truncation radius is small at the outburst peak, perhaps, smaller than $10R_{\rm S}$ 
(when we see no synchrotron emission), increases until 2007, when we see it at maximum, and decreases again in 2008.
The evolution of timing properties is consistent, in this scenario, with the spectral evolution, see Fig.~\ref{fig:spectra}.
The total X-ray flux drops and the spectrum hardens from 2005 Jul 6 to 2007 June 13, but then a re-brightening and a somewhat softer spectrum appears on 2008 Aug 10.
Moreover, a scenario of increasing truncation radius and development of the hot accretion flow in SWIFT~J1753.5--0127 is supported by broadband (optical to X-ray) 
spectral modelling \citep[fig.~8 of][]{KVT16}.

The evolution of the UV/optical fluxes during the first months of the outburst are also peculiar.
While the X-ray flux dropped by a factor of 15, the fluxes in the UV
range and in the B-filter drop by a factor of $\sim2$.  This behaviour was shown to be in conflict with 
reprocessing scenario \citep{CDSG10}.
This either suggests a dramatic change of reprocessing properties
(such as change of albedo) or is a signature of the appearance of an additional 
component (likely synchrotron radiation locally produced in the hot flow).
The latter explanation is more favoured by our findings.

We note that the disc to synchrotron ratio, $r_{\rm ds}$, has a low value during the decline stage, when the hot flow appears. 
The ratio then increases in the outburst tail, indicating the synchrotron flux is higher, relative to the disc, in the decline stage than in the tail. 
This trend is explained by the different scaling with the X-ray luminosity. 
For the long-term variations, such as the characteristic timescales of the outburst, the synchrotron scales roughly as $L_{\rm syn}\propto L_{\rm X}$ 
(as the spectral slope of the synchrotron self-Compton spectrum remains constant for different luminosities; see fig.~7b of \citealt*{VVP11}).
The observed disc spectrum likely falls in the transition between the Rayleigh--Jeans regime where $L_{\rm d} \propto T_{\rm d} \propto L_{\rm X}^{1/4}$
(where $T_{\rm d}$ is the irradiated disc temperature) and the blackbody peak with $L_{\rm d} \propto L_{\rm X}$, so we can roughly approximate 
$L_{\rm d} \propto L_{\rm X}^{1/2}$.
The ratio $\displaystyle L_{\rm d}/L_{\rm s}\propto L_{\rm X}^{-1/2}$ increases as the X-ray luminosity decreases.
A similar trend was observed in the BH binary XTE~J1550--564, the spectral evolution of which allowed the contributions of these two 
components to be separated \citep*{PVR14}.
Hence, the results of our timing analysis are in agreement with expectations from spectral studies.

% (Fig. 11)

\section{Conclusions}\label{sect:conclus}

For the first time we model the complex evolution of the
optical/X-ray timing properties of the 2005 outburst of the BH binary SWIFT~J1753.5--0127.
The shape of the optical/X-ray cross-correlation function demonstrates a single peak at the time of the outburst peak, which is then replaced 
by multiple dips and peaks during the outburst decline and a single dip plus peak structure in the outburst tail (see Fig.~\ref{fig:ccf_psd}).
We interpret this behaviour as the appearance of synchrotron emission from the hot flow towards the end of the outburst.
This scenario is also supported by the previously reported evolution
of the spectral properties, with disc Comptonization dominating the X-ray spectrum at the 
peak and synchrotron Comptonization dominating in the tail.

We propose a quantitative model describing the characteristics of each stage.
For the first time, the model is capable of explaining the CCF through the decline.
We show that coupling with the QPO significantly alters the shape of CCF.
The tentative presence of the QPO harmonic can be seen in the X-rays, however, its significance is very low.
We find that a second QPO harmonic is needed to account for the phase
lags at double the QPO frequency.
Similar to the first harmonic, X-ray and optical QPOs are
intrinsically connected and the phase shift is zero for both harmonics.

To describe the complex phase lags in the decline stage, two components in X-rays are needed: one coming from the disc- and the other from synchrotron Comptonization.
Interestingly, such a two-component X-ray light-curve naturally gives the humps in the X-ray power spectrum, which had to be modelled with additional Lorentzian components
or QPOs in previous works.

We note that the optical/X-ray cross spectra show excess power at high frequencies.
We attribute this feature to the limitations of the standard procedure of filtering out the noise by averaging over the data segments, which assumes that the statistical characteristics 
are time-invariant. 
We show that the trend of artificial power increase at high frequencies can be reproduced by simulating a number of CCFs with varying parameters.

We find that the CCF shape is an independent indicator of the accretion geometry.
We show that the study of correlated optical/X-ray variability allows us to probe radiative mechanisms operating in the immediate vicinity of the black hole. 
In the future, systematic study of the changing CCF shape during the outbursts of X-ray binaries may finally shed light on the physical processes
accompanying state transitions.

\section*{Acknowledgements}

AV thanks A. Zdziarski for the comments on the manuscript.
The work was supported by the Academy of Finland grant 268740 (AV, JP), the Foundations' Professor Pool and the Finnish Cultural Foundation (JP).
JJEK acknowledges support from the ESA research fellowship programme. MGR and SST were supported by the Russian Science Foundation grant 14-12-01287.

%\bibliographystyle{mnras}
%\bibliography{allbib}
%\end{document}

\end{document}